\documentclass[structabstract]{aa}
\usepackage{graphicx}
\usepackage{color}
\usepackage[normalem]{ulem}
\begin{document}

\title{CURious Variables Experiment (CURVE): \\
       CCD photometry of active dwarf nova DI Ursae Majoris}
\author{A. Rutkowski\inst{1}, A. Olech\inst{1}, M. Wi\'sniewski\inst{1}, P. Pietrukowicz\inst{1,2}, J. Pala\inst{3} 
       \and R. Poleski\inst{4}}
\institute{ Nicolaus Copernicus Astronomical Center,
           Polish Academy of Sciences, \\
           ul.~Bartycka~18, 00-716~Warszawa, Poland
           \email{(rudy,olech,mwisniew,pietruk)@camk.edu.pl}
       \and
           Departamento de Astronomia y Astrofisica, Pontificia Universidad 
Catolica de Chile, Casilla 306, Santiago 22, Chile
       \and
           J. Pala's Amateur Observatory, S{\l}upsk, Poland 
       \and
            Warsaw University Observatory, Al. Ujazdowskie 4, 00-478 Warsaw,
            Poland
           }

\abstract{We report an analysis of photometric behaviour of DI UMa,
an extremely active
dwarf nova. The observational campaign (completed in 2007) 
covers five superoutbursts and
four normal outbursts.}
{We examined principal parameters of the system 
to understand peculiarities of DI UMa, and other active
cataclysmic variables.}
{Based on precise photometric measurements, 
temporal light curve behaviour, $O-C$ analysis, and power spectrum analysis, we
investigated physical parameters of the system.} 
{We found that the period of the supercycle now equals 31.45$\pm$0.3 days. Observations during superoutbursts
infer that the period of superhumps equals $P_{\rm
sh} = 0.055318(11)$ days ($79.66\pm 0.02$ min). During quiescence, the light curve reveals a modulation of period 
$P_{\rm orb} = 0.054579(6)$ days ($78.59 \pm 0.01$ min), which we
interpret as the orbital period of the binary system.
The values obtained allowed us to determine a fractional period excess of
$1.35\% \pm 0.02\%$, which is surprisingly small compared to the usual value
for dwarf novae
(2\%-5\%). A detailed $O-C$ analysis was performed for two superoutbursts
with the most comprehensive coverage. In both cases, we detected an increase in
the superhump
period with a mean rate of $\dot P/P_{\rm sh} = 4.4(1.0)\times 10^{-5}$.}
{Based on these measurements, we confirm
that DI UMa is probably a \emph{period bouncer},  
an old system that reached its period minimum a long time ago, has a secondary
that became a degenerate brown dwarf,  the entire system evolving now toward
longer periods. DI UMa is an extremely interesting object
because we know only one more active ER UMa star with similar characteristics
(IX Dra).}

\authorrunning{Rutkowski et al.}
\titlerunning{CCD Photometry of DI UMa.}
\maketitle

\section{Introduction}

The ER~UMa type stars (Kato and Kunjaya
\cite{kato1995})  remains one of the most intriguing systems among all
cataclysmic variables. Those objects, belonging to SU UMa-type
dwarf novae, exhibit various types of behaviour, such as normal outburst,
superoutburst, and superhumps. Smak (\cite{smak1984}) identified the
origin of normal outbursts in dwarf novae in terms of thermal instability of
the accretion disc. Whitehurst (\cite{whitehurst1988}) and Osaki
(\cite{osaki1989}) then proposed a model that  described superoutburst phenomena
in those stars (for a review see Osaki \cite{osaki2005}).
The period between two consecutive superoutbursts (also called
the supercycle period)  in ER~UMa type objects is about a few tens of days.
Until now, the shortest known supercycle for this class of stars has been
$P_{so}=19.07$ days, the value measured for RZ~LMi (Robertson et al.
\cite{robertson1995}, Olech et al. \cite{olech2008}). 

DI UMa is a member of ER~UMa class. The history of
observations of DI~UMa starts in 1959, when Hoffmeister (\cite{hoffmeister1959})
discovered a rapidly varying star, which was later designated as DI~UMa type.
Further spectroscopical studies identified this object as a cataclysmic
variable (Bond \cite{bond1978}, Szkody \& Howell \cite{szkody1992}).
Kato, Nogami \& Baba (\cite{kato1996}) determined the period of
the supercycle to be 25 days. Soon after, Fried et al. (\cite{fried1999}) found
that superoutbursts a have recurrence time of 30-45 days, and the superhump
period and the orbital period  equal 0.05529(5) and 0.054564(1) days,
respectively. Hence, they measured the superhump period excess of
$1.3\%$,  which was then one of the lowest for known CVs.
Although a small deviation from regularity in the occurrence of
superoutburst is admissible, this prominent change in supercycle length
remains problematic from the standard thermal--tidal instability model
point of view. This fact, in addition to a low period excess, that is appropriate for quiet WZ Sge stars, both with global properties and
parameters characterizing this system, have encouraged us to take a
closer look at the behaviour of DI~UMa.

\section{Observations and data reduction}

Observations were carried out between January 15 and June 25 in 2007.
Two telescopes were used to collect the data.
The first was a 0.6-m Cassegrain
telescope of Warsaw University Observatory located at its Ostrowik station.
It is equipped with a Tektronix TK512CB
back-illuminated CCD camera. The image scale was $0''.76/$pixel
providing a $6'.5 \times 6'.5$ field of view (Udalski and Pych 
\cite{udalski1992}). The second was a f/3.8 0.3-m Newton telescope located at Jacek Pala's
amateur observatory in S\l{}upsk, equipped with a SBIG ST-2K dual CCD
camera.

Observations of DI UMa reported in this paper cover 36 nights and
include 3093 measurements. The average exposure time was about 150 sec.
Although we have not used an autoguider, images of stars remain unshifted
and PSF profiles seem undistorted.

For the observational data processing, we used the IRAF\footnote{ IRAF is
distributed by the National Optical Astronomy Observatory, which is
operated by the Association of Universities for Research in Astronomy,
Inc., under cooperative agreement with the National Science Foundation.}
package. The profile photometry was then obtained using the DAOphotII
package (Stetson \cite{stetson1987}). A standard way of reduction was
performed for the Ostrowik data. Unfortunately, amateur observations
from S{\l}upsk were obtained without flat--field calibration images, so
only simplified reduction was possible. This measurements are of poorer
accuracy but sufficient to be included in the analysis.

Observations were carried out without filters for two reasons. First, due to
the lack of
an autoguiding system, we wished to maintain exposures as short as possible to
minimize guiding errors. Second, because our main goal was an analysis of
the temporal behaviour of the light curve, the use of filters could
cause the object to become too faint to be observed in quiescence.

Relative unfiltered magnitudes of DI UMa were determined to be the
difference between the magnitude of the variable and the magnitude of a
nearby comparison star. Although we do not include AAVSO measurements
in further analysis, we used these data to draw conclusions (from
the overlap of two light curves) about the mean $V$ magnitude of the
star. This procedure may introduce a relatively high systematic
uncertainty in transforming the magnitudes, probably even as large
as $\sim$ 0.3 mag in the zero point.

Table 1 presents a journal of our CCD observations of DI UMa. 
In the following parts of this article (for convinience), civil
dates are used to
characterize  moments of observations, rather than the corresponding HJD.
In total, we observed the star for almost 113 hours \footnote{The photometric
data discussed in this paper are available in electronic form at the CDS via 
anonymous ftp to cdsarc.u-strasbg.fr (130.79.128.5) or via 
http://cdsweb.u-strasbg.fr/cgi-bin/qcat?J/A+A/}.

\begin{table}[h!]
\caption{Observational journal for the DI UMa campaign.}
\centering
\smallskip
{\small
\begin{tabular}{l c c c c c} \hline \hline
Date in & Start [HJD]& End [HJD]& Dur & No. of & $<V>$ \\
2007 & (2454000+) & (2454000+) & (hr) & points & (mag) \\
\hline
Jan 15 & 116.40818 & 116.64998 & 5.80 & 101 & 14.84 \\
Mar 07 & 167.33473 & 167.56941 & 5.63 & 97 & 16.23 \\
Mar 13 & 173.49266 & 173.65707 & 3.95 & 117 & 15.26 \\
Mar 26 & 186.33092 & 186.64037 & 7.43 & 99 & 17.69 \\
Mar 27 & 187.32404 & 187.52248 & 4.76 & 89 & 16.95 \\
Mar 28 & 188.33963 & 188.55087 & 5.07 & 106 & 15.71 \\
Mar 30 & 190.35326 & 190.46698 & 2.73 & 48 & 17.51 \\
Mar 31 & 191.36005 & 191.38821 & 0.68 & 9 & 17.69 \\
Apr 04 & 195.30920 & 195.31517 & 0.14 & 7 & 17.41 \\
Apr 05 & 196.33060 & 196.35222 & 0.52 & 24 & 15.73 \\
Apr 11 & 202.31427 & 202.34950 & 0.85 & 26 & 17.65 \\
Apr 12 & 203.34770 & 203.46420 & 2.80 & 79 & 17.46 \\
Apr 13 & 204.34525 & 204.47620 & 3.14 & 96 & 15.98 \\
Apr 14 & 205.36642 & 205.52808 & 3.88 & 115 & 14.70 \\
Apr 15 & 206.27831 & 206.47507 & 4.72 & 172 & 14.67 \\
Apr 16 & 207.28708 & 207.58366 & 7.12 & 523 & 14.82 \\
Apr 17 & 208.27757 & 208.40331 & 3.02 & 102 & 14.87 \\
Apr 18 & 209.34435 & 209.53969 & 4.69 & 123 & 14.93 \\
Apr 20 & 211.35228 & 211.46002 & 2.59 & 69 & 15.06 \\
Apr 21 & 212.28355 & 212.37567 & 2.21 & 61 & 15.21 \\
Apr 23 & 214.37639 & 214.46617 & 2.15 & 58 & 15.52 \\
Apr 25 & 216.28540 & 216.52020 & 5.64 & 176 & 17.15 \\
Apr 26 & 217.32268 & 217.44514 & 2.94 & 38 & 17.40 \\
Apr 29 & 220.39641 & 220.51590 & 2.87 & 22 & 16.08 \\
May 16 & 237.44159 & 237.51015 & 1.65 & 42 & 14.77 \\
May 17 & 238.33788 & 238.49019 & 3.66 & 139 & 14.83 \\
May 18 & 239.33408 & 239.51366 & 4.31 & 136 & 14.94 \\
May 20 & 241.36450 & 241.40883 & 1.06 & 24 & 15.11 \\
May 21 & 242.36734 & 242.46793 & 2.41 & 63 & 15.22 \\
May 22 & 243.38513 & 243.47040 & 2.05 & 52 & 15.30 \\
May 23 & 244.37801 & 244.48001 & 2.45 & 63 & 15.30 \\
May 24 & 245.34680 & 245.49097 & 3.46 & 66 & 15.61 \\
May 25 & 246.33298 & 246.43703 & 2.50 & 26 & 16.66 \\
Jun 12 & 264.35449 & 264.36172 & 0.17 & 4 & 17.95 \\
Jun 19 & 271.37112 & 271.46188 & 2.18 & 53 & 14.72 \\
Jun 20 & 272.34845 & 272.41129 & 1.51 & 48 & 14.74 \\
\hline
\label{table1}
\end{tabular}}
\end{table}

\section{Global light curve}

\begin{figure*}
\centering
\includegraphics[angle=-90,width=\textwidth]{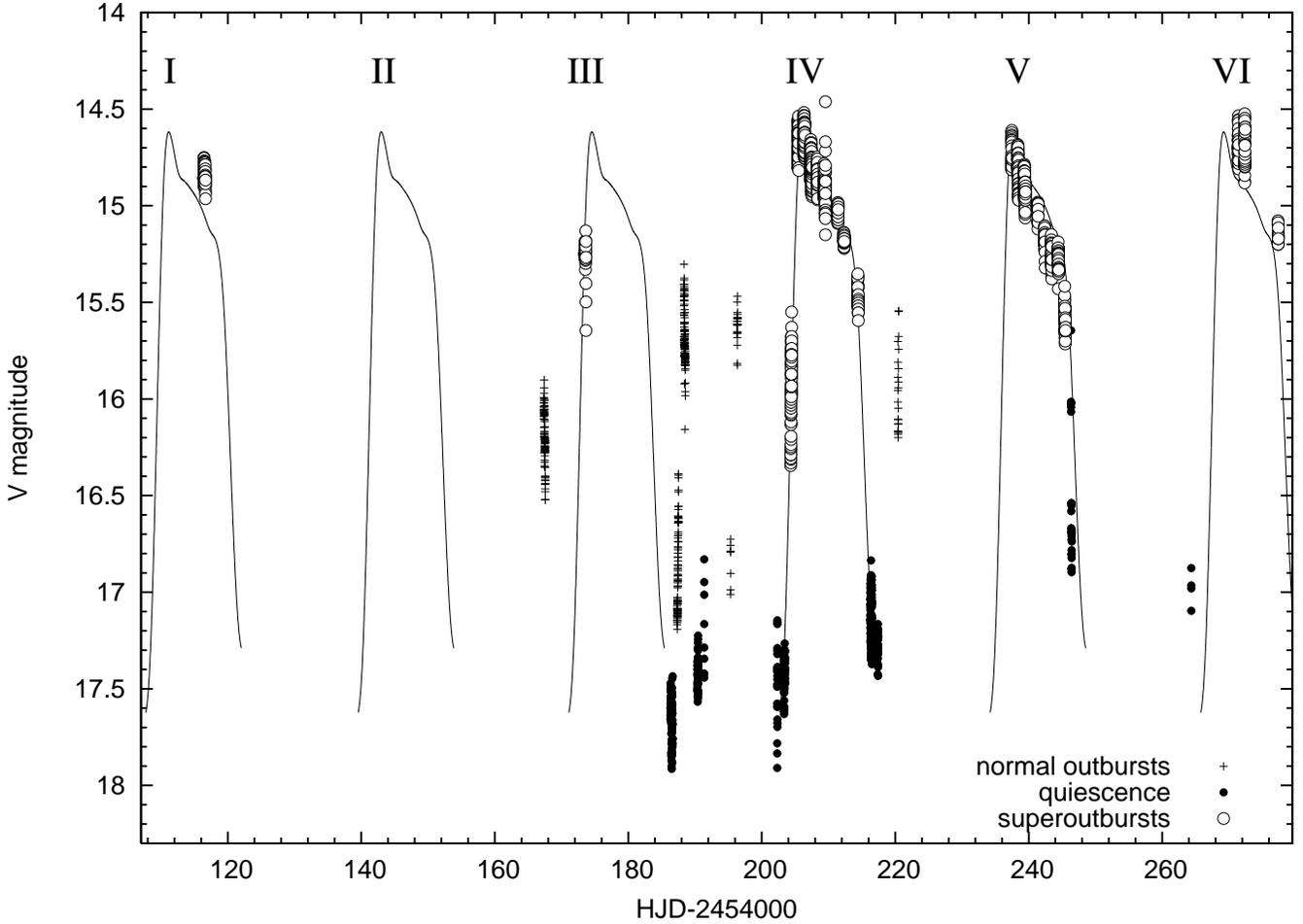}
\caption{ Global light curve of DI UMa during the 2007 campaign.
Dots, plus signs and circles indicate
different periods of DI UMa activity. Those symbols represents quiescent state,
normal outbursts and superoutbursts stage, respectively.  Such
division is used for further analysis (as in the text).
Solid line shows a fit of the superoutburst profile no. IV, which is repeated 
every 31.45 days backward and forward.}
\label{profile}
\end{figure*}

Figure \ref{profile} presents the photometric behaviour of DI UMa during
this campaign. The shape of the light curve corresponds to the standard
picture of active dwarf novae (Olech et al. \cite{olech2004b},  Kato and
Kunjaya \cite{kato1995},  Robertson et al. \cite{robertson1995}).
Frequent superoutbursts reach $V\approx14.5$ magnitude at maximum
and fade to $V \approx 17.8$ at quiescence, so the amplitude of the
superoutburst reaches $A_{sup} \approx 3.3$ mag. Between  prominent
superoutbursts, we can identify normal outbursts, reaching $\approx 15.4$ mag
with an amplitude of $\approx 2.4$ mag. 

The predictions of typical scenarios developed to explain the dwarf--novae
light curve, in particular, its excess of brightness during superoutburst
(compared to normal outburst brightness) as well as its duration time and profile, are in agreement with our observations. This allowed us to
distinguish unambigously between measurements collected in different phases of activity.
From the global light curve, we then selected only those nights
during which the star was in superoutburst (marked by open circles).
For the resulting light curve, we then computed the power spectrum using
the ZUZA code of Schwarzenberg-Czerny (\cite{Schwarzenberg-Czerny1996}).

\begin{figure}
\centering
\includegraphics[angle=-90,width=0.5\textwidth]{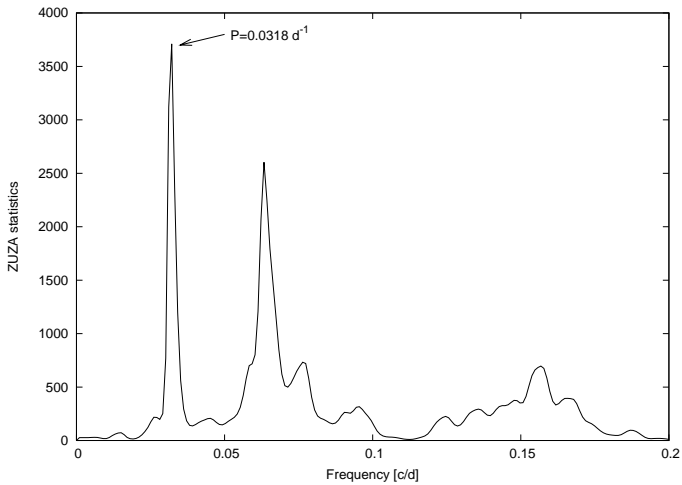}
\caption{ Power spectrum of DI UMa light curve after removal of the data
from quiescence and normal outbursts. The peak corresponding to the supercycle
period is marked by an arrow.}
\label{powershape}
\end{figure}
\begin{figure}
\centering
\includegraphics[angle=-90,width=0.5\textwidth]{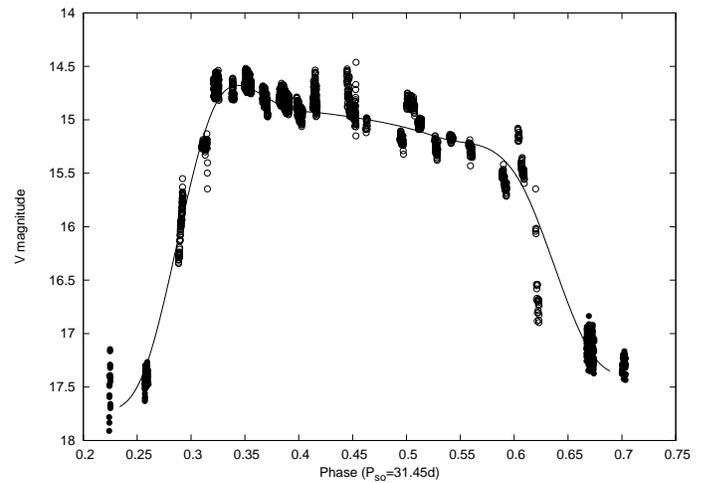}
\caption{ Light curve of DI UMa in superoutbursts obtained by folding
the general light curve with a supercycle period of 31.45 days. }
\label{shape}
\end{figure}

The power spectrum was computed for a range of speeds 0 - 0.2 c/d;  the
frequencies  found to represent the longest periods that could be
seen in the light curve and reflect the variability related to
superoutbursts.  The resulting periodogram is shown in Fig.~\ref{powershape}.
Two prominent peaks can be observed.  The highest peak
at frequency of 0.0318 c/d corresponds to the period of the
superoutburst cycle, which according to our measurements is
31.45 days. This value differs significantly  from previous
estimates made by Kato et al. (\cite{kato1996}) and agrees with the
range measured by Fried et al. (\cite{fried1999}).
Since, our light curve corresponds to 6 consecutive superoutbursts, the
observational coverage is of higher quality than in earlier measurements and
allowed us to achieve higher accuracies of period determinations.

As one can see from Fig. \ref{profile}, the profiles
of superoutbursts are fitted with solid lines.  Those best-fit solutions
were obtained for data belonging only to superoutburst IV, which had
sufficient observational coverage.  An average superoutburst profile was
obtained by completing an analytical fit using Bezier curves. 
The derived profile was copied, in both forward and backward directions, 
with supercycle period of 31.45 days (as determined before), taking into
account all acquired observational data. This line serves only as a check and
demonstration tool and is not used in later analysis,
as for example in the detrending
procedure described in Sect. \ref{perioda}.
During the observing season, the stability of the
supercycle was quite high, and the analytical light curve reproduce
observations  well. 

Figure \ref{shape} presents the phased light
curve folded with the supercycle period of $P_{so}=31.45$ days.
As one can see, the shape of the superoutburst is typical. First, the
brightness of the star increased rapidly from the 17.8 mag, during quiescent level,
to a peak magnitude of around 14.5 mag. This increasing phase lasted
$\sim 2$ days. A plateau with a slowly decreasing trend
of rate  0.8 mag/day, lasting $\sim 9.4$ days was observed. At the end
of the plateau, the star reached $\sim 15.7$ mag, then a far steeper
decline occurred, and the final decline lasted $\sim 4.1$ days. Based
on this figure, we conclude that on average the entire  superoutburst
lasts around $\approx 17.7$ days. In total, we have five certain
detections of the superoutburst and one more (no. II) added as a prediction.

\section{Superhumps}

\begin{figure*}
\centering
\includegraphics[angle=0,width=0.9\textwidth]{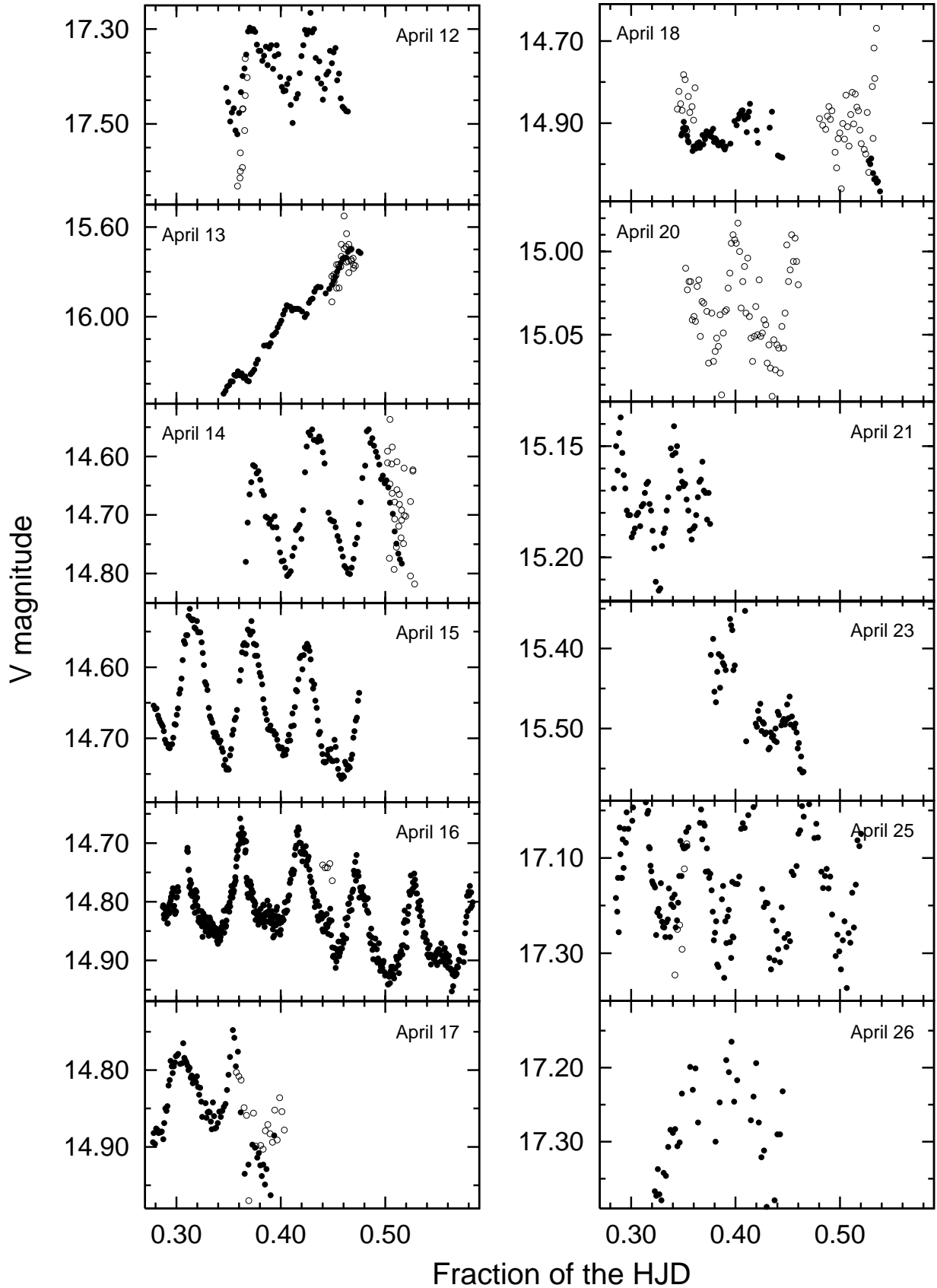}
\caption{ Nightly light curves of DI UMa from its April 2007 superoutburst.
Consecutive nights are denoted by dates in top-left/right corner. Dots shows
the data obtained in Warsaw University Observatory, whereas open circles denotes
observations made in S\l{}upsk Observatory.}
\label{super}
\end{figure*}

Figure \ref{super} shows the light curves of DI UMa during individual
nights of the superoutburst that occurred in April 2007 (marked by roman number IV
in Fig. \ref{profile}). Data from twelve almost consecutive nights
(gaps on Apr 19, Apr 22 and Apr 24) are
presented in this plot. One can clearly observe a variation in the superhumps
from orbital humps (April 12), a steep initial rise with a trace of modulation
(April 13), through fully-developed, tooth-shape, common superhumps (April
14-21), to gradual disappearance and, finally, a transition to quiescence
and orbital humps (April 25-26). At maximum light, the  amplitude of
clearly visible common superhumps reaches almost 0.3 mag.

\subsection{Periodicity analysis}\label{perioda}

As one may see from previous figures, the light curves of DI UMa are
characterized by an alternating  trend of increasing and
decreasing and so on.
Thus, the magnitudes had to be transformed to a
common $V$ system. The fact that the data from quiescence to
maximum of superoutbursts exhibited a variability in the range of about 3.3 mag,
can have a pernicious effect on the period search. Taking this into acount, we
prepared the data by subtracting the mean (and trend if necessary) from each
 night of time-series data as follows.

The data from each night were fitted with
a straight line or a parabola and this fit was subtracted from the true
light curve. This allowed us to remove the trend in all nightly light curves
and shift the data from the entire campaign to a common level. As a result,
the data have average brightness equal to zero and consist of only
short-term modulations with periods significantly shorter than one
day. After this approach, we expect an improvement in the reliability of
the period determination.

One can expect that the power spectrum based on all available data can be
contaminated by several smaller peaks and aliases  due to the presence of various periodicities in the light curve.  Thus, based only on
one periodogram covering all available data, it is difficult to
extract reliably all frequencies present in the light
curve. For this reason, we decided to perform a more detailed frequency
analysis.

\subsubsection{Superoutburst data analysis} \label{sda}

We consider again Fig \ref{profile}. We separated all points into
three types of data. Points marked by open circles are measurements
completed during the superoutburst phase. These points, after the removal of 
trends,  were considered first in the analysis. The periodicity search
was completed using the ZUZA code (Schwarzenberg-Czerny
\cite{Schwarzenberg-Czerny1996}).

\begin{figure}[h!]
\begin{center}
\includegraphics[angle=0,width=0.5\textwidth]{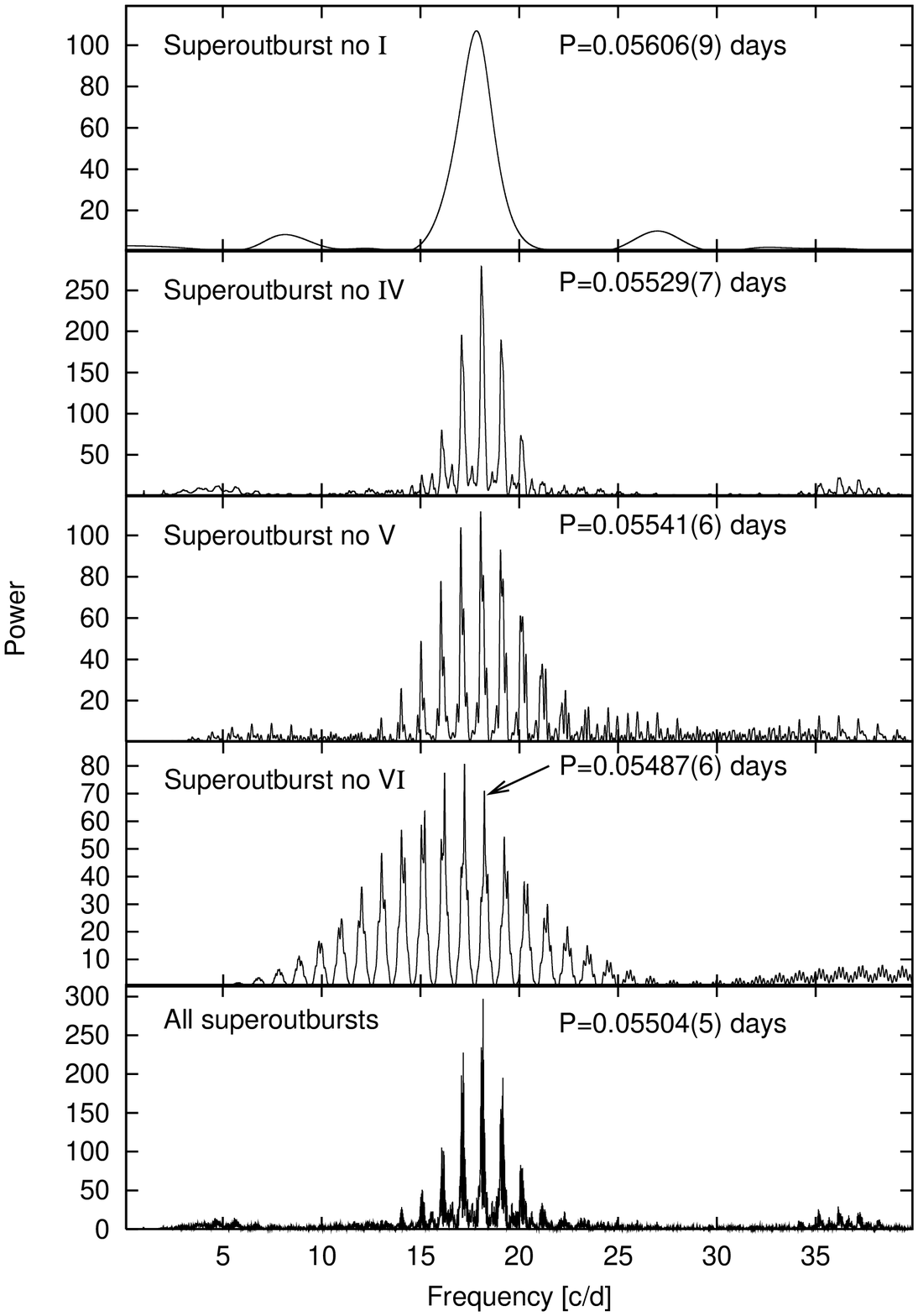}
\caption{ Power spectra for superhumps observed during four 
superoutbursts of DI UMa. The numbers near the most prominent peaks 
represent the most probable value of superhump period.}
\label{combine}
\end{center}
\end{figure}

The lowest panel of Fig. \ref{combine} presents the resulting
periodogram. The most prominent peak correspond to the superhump period
equal to $P_{\rm sh}=0.05504(5)$ days. This value is in agreement with
the previous determination completed by Fried et al. (\cite{fried1999}).

We also checked the period stability of superhumps during
successive superoutbursts. Figure \ref{combine} shows a comparison of
periodograms obtained for different superoutbursts. As one can see,
there is no sign of a significant superhump period change
from one superoutburst to another. Nevertheless, periodograms for
different superoutbursts may be inconclusive due to insufficient
data being available used in their preparation. There is only one
night for the superoutburst no. I, and three nights for the superoutburst
no. VI. For the same reason, aliases can distort significantly the
obtained spectrum.

\begin{figure}
\begin{center}
\includegraphics[angle=-90,width=0.5\textwidth]{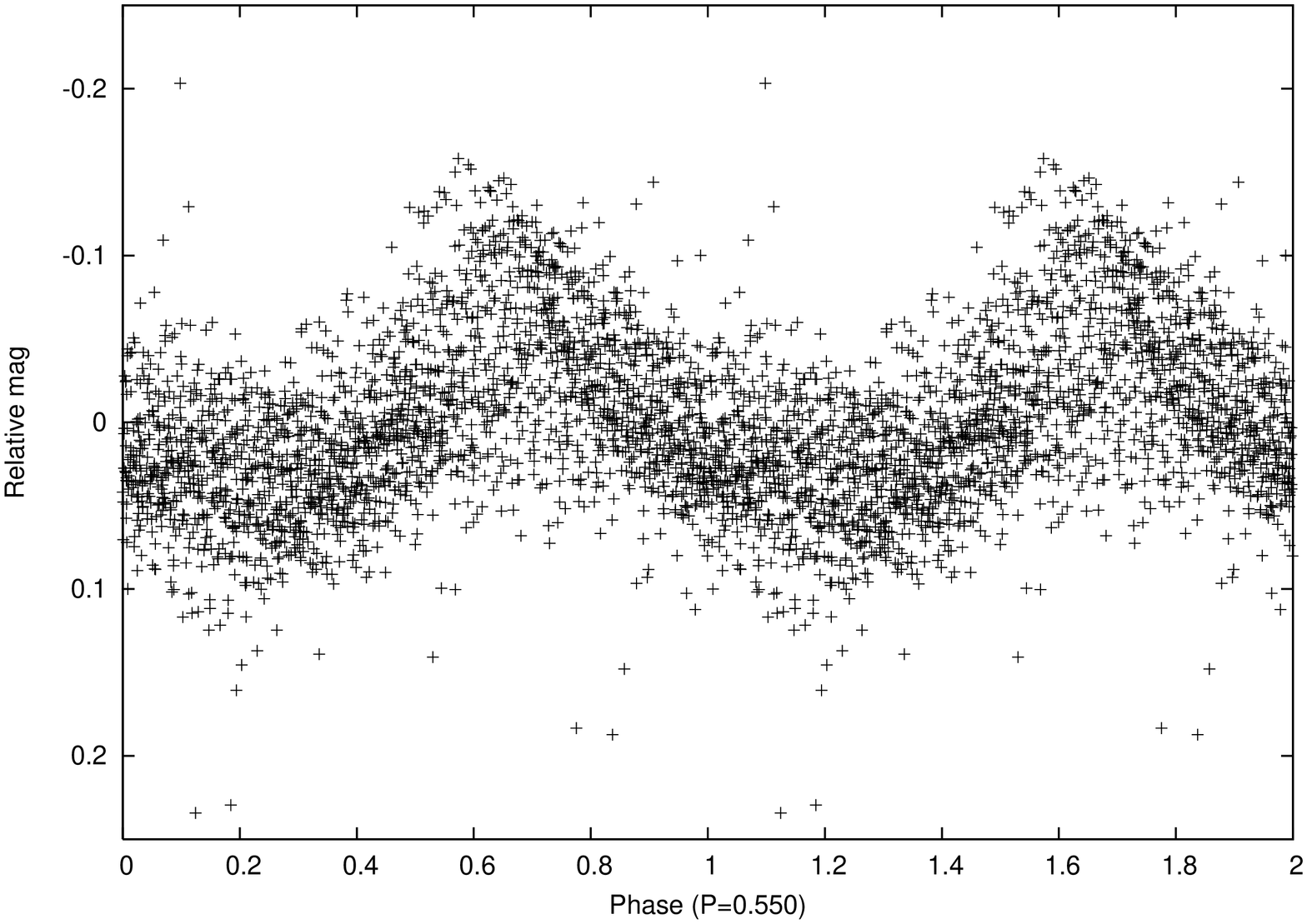}
\caption{ Detrended light curve from data collected during all observed
superoutbursts in 2007 campaign, folded with superhump period
of $P_{sh}=0.05504$ days.}
\label{phase_out}
\end{center}
\end{figure}

The light curve from which trends have been substracted that contain
 superhumps from all superoutbursts 
might be out of phase by an entire period exhibiting no significant evidence of phase shift between superhumps from different superoutbursts. Such a light
curve phased with a period $P=0.05504$ days is presented in Fig.
\ref{phase_out}.

\begin{figure}
\begin{center}
\includegraphics[angle=0,width=0.5\textwidth]{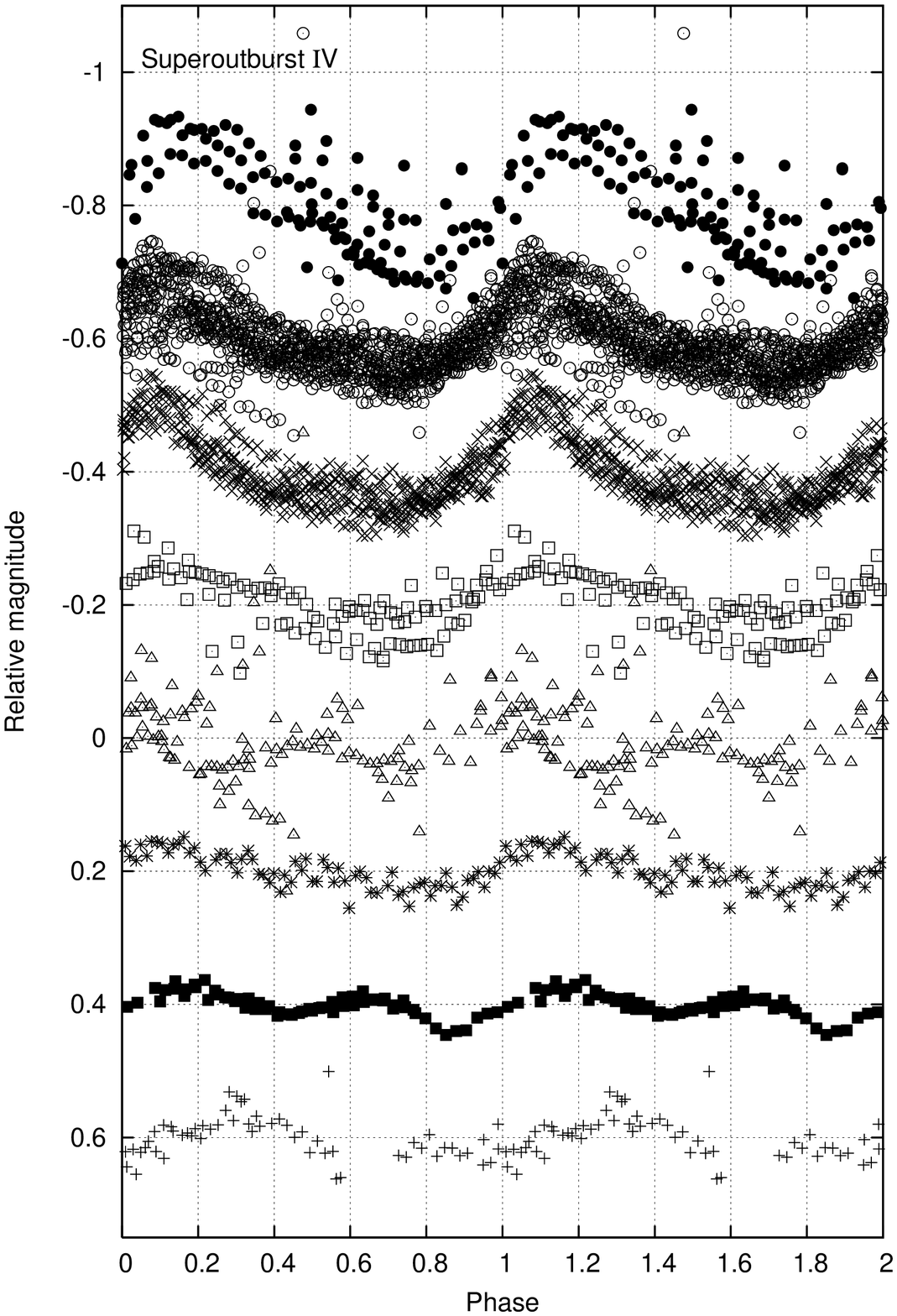}
\caption{ Stability of superhumps phase during
eight nights covering the April 2007 superoutburst, from April 14 (top to
April 23 (bottom). Light curves are phased with superhump 
period $P_{\rm sh}=0.05504$ days.}
\label{shift}
\end{center}
\end{figure}

To check the stability of the phase of superhumps, we
plot phased light curves separately for each night. We selected the
superoutburst no. IV, for which we have data of the highest quaility
coverage. Figure \ref{shift}
presents the result of this approach. Successively, light curves
from April 14 to 23 are shown from top to bottom in the plot.
During the first four nights only, a slight or no phase-shift was observed.
Later on (around April 18), one can see the emergence of a secondary hump,
although with out significant phase reversal, in contrast to some ER UMa 
stars (for example see Kato et al. \cite{kato2003}).
%
%
One may expect that a modulation connected with the orbital period, present
in a quiescent light curve, may also be detected in superoutbursts as 
in the case of another active dwarf nova, IX Dra (Olech et al. \cite{olech2004a}). This idea
encouraged us to search for frequencies correlated with orbital humps in the
power spectrum. We decided to perform prewhitening of the detrended
light curve of DI UMa from all superoutbursts. First we removed
from the light curve a modulation with a period corresponding to
superhumps. We then performed a power spectrum analysis on the
obtained data. We derived a noisy power spectrum with no
significant peak, indicating that orbital modulation was absent during
the superoutburst or its amplitude was below our detection limit.

\subsubsection{Quiescence data analysis} \label{qda}

\begin{figure}
\begin{center}
\includegraphics[angle=0,width=0.5\textwidth]{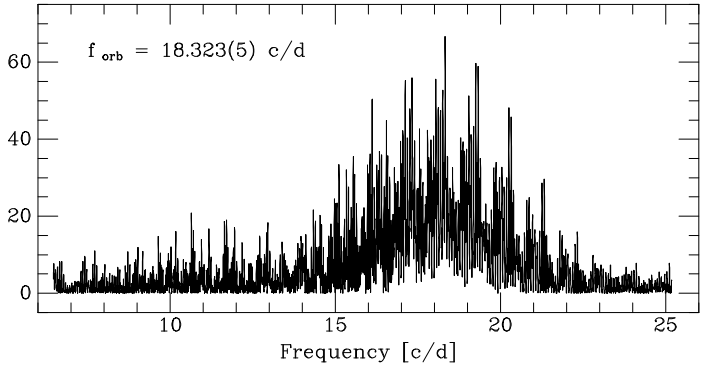}
\caption{ Power spectrum of data collected during quiescence.} \label{quiet}
\end{center}
\end{figure}

\begin{figure}
\begin{center}
\includegraphics[angle=-90,width=0.5\textwidth]{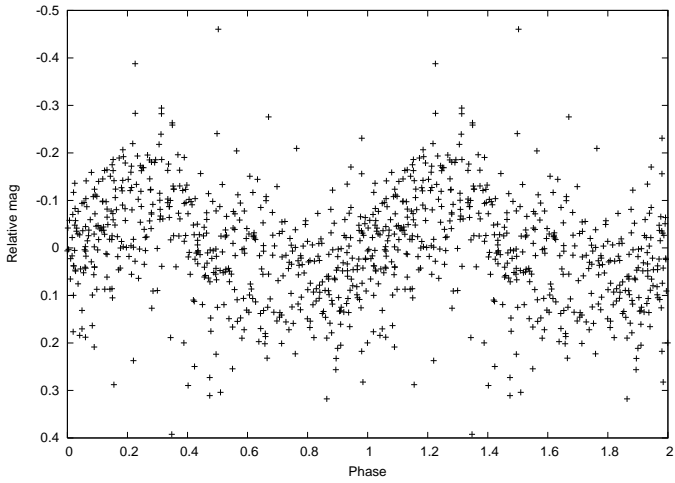}
\caption{ Detrended light curve of data collected during quiescence and 
phased with period $P_{\rm orb}=0.05458$ days, which we interpret as the orbital 
period of the system.} \label{quiet1}
\end{center}
\end{figure}

We now consider the analysis of the data collected during quiescence (black dots
in Fig. \ref{profile}).
We repeated the
approach from Sect. \ref{sda} and produced a periodogram using the ZUZA
software. Figure \ref{quiet} presents the results of this procedure. One of the most
prominent peaks present in the power spectrum is located at a frequency 
of $f_{\rm orb}=18.323\pm0.005$~c/d ($P_{\rm orb}=0.054576(15)$ days). From a statistical point of view, it is possible
that some other, lower, peak represents the true modulation present in the
quiescent light curve of DI UMa. However we recall the result of Fried
et al. (\cite{fried1999}), who obtained photometrically the orbital period of DI UMa equal to
$P_{\rm orb}=0.054564(1)$ days, which is in excellent agreement with our
determination. Hence, we assume the value of $P_{\rm orb}=0.054576(15)$ days
as the most probable for the orbital period of the binary. We phased our quiescence data based on this determination. The result of this procedure is
shown in Fig. \ref{quiet1}. 

In addition, we  collected data during normal outbursts, but
due to an insufficient amount of measurements we were unable to derive useful
conclusions for these data.

\subsection{The O-C analysis}

DI UMa often exhibits clear and periodic modulation in quiescence, with
an amplitude reaching even 0.4 mag. For this modulation, we determined 15
moments of maxima. A least-squares linear fit to these data gives the
following ephemeris:

\begin{equation}
\textrm{HJD}_\textrm{orb-max} = 2454186.341(4)+0.054580(7) \times E,
\end{equation}

\noindent which agrees within the errors with the determination based on the
power spectrum described in the previous paragraph.  The combination of
both of these measurements gives us our final value of the orbital period of
the binary, which is equal to $P_{\rm orb} = 0.054579(6)$ days ($78.59 \pm 0.01$ min).

On the other hand, the light curve of DI UMa from superoutbursts contains 38
moments of maxima.

The observations from superoutbursts no. IV and V contain sufficient data to
perform a detailed analysis and draw valuable conclusions about the temporal
evolution of the superhump period.

\begin{figure}[h!]
\begin{center}
\includegraphics[angle=0,width=0.45\textwidth]{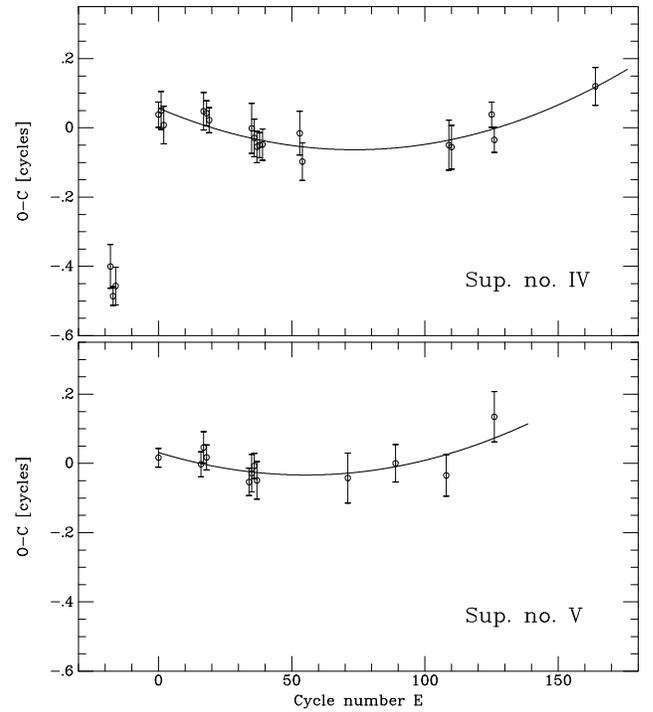}
\caption{ $O-C$ diagram for superhumps of DI UMa occurring during the
April 2007 (IV) and May 2007 (V) superoutbursts.
Solid lines correspond to quadratic ephemeris. Points with
negative cycle numbers might be connected with orbital hump and not with
superhumps.}
\label{oc}
\end{center}
\end{figure}

The maxima from superoutburst no. IV might be fitted with a common ephemeris in
the form:

\begin{equation}
\textrm{HJD}_\textrm{max} = 2454205.3729(9)+0.055320(12) \times E
\end{equation}
The $O-C$ values computed according to this ephemeris are shown in the top panel
of Figure \ref{oc}. It is clear that they show a slightly increasing trend and thus
we decided to fit the maxima with a quadratic ephemeris in the form:
\begin{eqnarray}
\textrm{HJD}_\textrm{max} = 2454205.3761(12)+0.055139(50) \times E 
\nonumber \\
+ 1.23(33)\cdot 10^{-6}\times E^2
\end{eqnarray}
We emphasize that the maxima denoted by negative cycle numbers
relate to  the night of April 12, i.e. from the rapid initial rise to the
superoutburst, and were not included in the above fits. They are shifted in
phase by about half a cycle and might not correspond to ordinary
superhumps but to early superhumps or even orbital humps.
The light curve for April 13 resembles the shape of early superhumps 
 known from other studies (Kato \cite{kato2002},
Maehara \cite{maehara2007}). Osaki \& Meyer (\cite{osakimeyer2002})
proposed that early superhumps were
caused by two-armed dissipation pattern on the acretion disc.
During our observations, the early superhumps have a
double-peaked profile and its amplitude was in the range of 0.05-0.07 mag.
Unfortunately, since the distinctive shape of early sperhumps was
visible only in April 13, the precise determination of its period was
imposible. The crude estimate (read from graph) infers that
$P_{esh}\cong0.0540$ days with high uncertainty about 1\%. In light of previous
observations of early superhumps (for example Patterson \cite{patterson2003}),  suggesting that $P_{esh}$  is equal to around orbital period,  we consider this value with caution.

The maxima from superoutburst no. V can be fitted with a linear ephemeris
of the form:
\begin{equation}
\textrm{HJD}_\textrm{max} = 2454237.46511(97)+0.055313(22) \times E
\end{equation}
In this case, a slightly increasing trend in the superhump period
was also observed, and the maxima could be fitted with a quadratic ephemeris,
which is given by the following equation:
\begin{eqnarray}
\textrm{HJD}_\textrm{max} = 2454237.4669(13)+0.055183(68) \times E
\nonumber \\
+ 1.18(58)\cdot 10^{-6}\times E^2
\end{eqnarray}
The superhump periods inferred from linear fits can be used together to compute
our final  mean value of the superhump period, which is
$P_{\rm sh} = 0.055318(11)$ days ($79.66\pm 0.02$ min).

We note that period derivatives obtained during both
superoutbursts are consistent within errors and provide a relatively
low value of $\dot P/P_{\rm sh} = 4.4(1.0)\times 10^{-5}$.

\begin{figure}[h!]
\begin{center}
\includegraphics[angle=-90,width=0.52\textwidth]{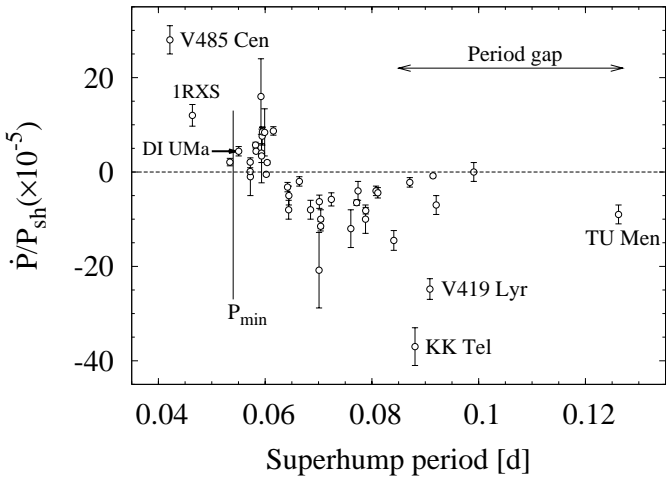}
\caption{Relation between the superhump period and its derivative for known
SU UMa stars.}
\label{pdot}
\end{center}
\end{figure}

We  recall the figure from Uemura et al. (\cite{ue2005}), which was updated
by Rutkowski et al. (\cite{rut2007}), showing the relation between the period
derivative and the superhump period. It is presented in Fig. \ref{pdot}.
The figure shows that DI UMa is placed in the group of small superhump periods
($P_{sh}$) and exhibits a low but positive $P/P_{sh}$ value.

\section{Conclusions}

Our five-month observational campaign concerning  an active SU UMa-type star
DI~UMa resulted in the detection of five superoutbursts,  which repeat
every 31.45 days. This is far longer than  the 25
days obtained by Kato, Nogami \& Baba (\cite{kato1996}) but falls within
the range of 30--45 days observed by Fried et al. (\cite{fried1999}).

Our observations indicate that DI UMa exhibits periodic, light modulations
both in quiescence and during superoutbursts. The first of these modulations, with 
$P_{\rm orb} = 0.054579(6)$ days ($78.59 \pm 0.01$ min), is interpreted
as the orbital period of the system and the second, with  $P_{\rm
sh} = 0.055318(11)$ days ($79.66\pm 0.02$ min), as the superhump period.

A very orbital period and a small period excess equal to only $1.35\%
\pm 0.02\%$ suggest that DI UMa is a so-called \textit{period bouncer}, i.e. an old system that reached its period minimum a long time ago, its secondary
became a degenerated brown dwarf and the entire system now evolves toward
longer periods (Patterson \cite{patterson2001}). DI UMa is thus unique because we know
only one more active ER UMa star with similar characteristics - namely IX
Dra (Olech et al. \cite{olech2004a}).

Inspection of Fig. \ref{pdot} indicates that long-period systems tend to show
large negative values of $\dot P/P_{\rm sh}$, whereas short-period systems 
are characterized by small and positive period derivatives.
This picture has a physical interpretation proposed
by Osaki and Meyer (\cite{osaki2003}). In long-period dwarf-novae with high 
transfer rates, the 3:1 resonance radius in the accretion disc is close
to the tidal truncation radius, and  eccentric waves in the disc
may thus propagate only inwards, proding a decrease in the superhump period. 
In short-period systems with lower accretion rates and more infrequent
outbursts, the 3:1 radius is much smaller than the tidal truncation radius
and sufficient matter is stored in the disc to cause the propagation of outward eccentric waves. In these systems, the superhump period may increase.

Do the properties of DI UMa inferred from observational data agree with this
scenario? It is a short-period but active dwarf nova,
showing both frequent outbursts and superoutbursts, which are indicate of a 
high-mass transfer-rate. It may imply that the disc contains sufficient matter
to reach not only the 3:1 resonance radius, but also the region of the 2:1 
resonance radius, as
in the case of IX Dra, another active but old dwarf nova. There is thus a possibility of the ignition of outwardly propagating eccentric waves, which may cause
an increase in the superhump period.

We recall that DI UMa is the second shortest-period 
dwarf nova\footnote{There are two variables, V485 Cen (Olech
\cite{olech1997}) and EI Psc (Uemura et al. \cite{ue2002}), which show far
 shorter periods
of superhumps but their status as classical SU UMa systems is unknown.}
with a determined period derivative. Smaller values of both the superhump
period and its derivative were determined for VS 0329+1250 (Shafter et al. \cite{Sch2007}).

\begin{acknowledgements}
We acknowledge generous
allocation of  the Warsaw University Observatory 0.6-m telescope time. This work
used the on-line service of the VSNET and AAVSO. We would like to thank
Prof. J\'ozef Smak for fruitful discussions. We also thank Maciej Bilicki
for carefully reading this manuscript. This work was supported by
MNiSzW grant no. N N203 301335 to A.O.
\end{acknowledgements}

\end{document}